\documentclass[a4paper,11pt]{article}

\usepackage{jinstpub} 
\usepackage{epsfig}

\title{\boldmath The Experimental Program for High Pressure Gas Filled Radio Frequency Cavities for Muon Cooling Channels}

\author[a,1]{B. Freemire,\note{Corresponding author.}}
\author[b]{M. Chung,}
\author[c]{P.M. Hanlet,}
\author[d]{R.P. Johnson,}
\author[e]{A. Moretti,}
\author[f]{Y. Torun,}
\author[e]{K. Yonehara}

\affiliation[a]{Northern Illinois University, Illinois, USA}
\affiliation[b]{Ulsan National Institute of Science and Technology, Ulsan, South Korea}
\affiliation[c]{Tokamak Energy, Milton Park, Oxfordshire, UK}
\affiliation[d]{Muons, Inc., Batavia, Illinois, USA}
\affiliation[e]{Fermi National Accelerator Laboratory, Batavia, Illinois, USA}
\affiliation[f]{Illinois Institute of Technology, Chicago, Illinois, USA}

\emailAdd{bfreemire@niu.edu}

\abstract{An intense beam of muons is needed to provide a luminosity on the order of 10$^{34}$ cm$^{-2}$s$^{-1}$ for a multi-TeV collider.  Because muons produced by colliding a multi-MW proton beam with a target made of carbon or mercury have a large phase space, significant six dimensional cooling is required.  Through ionization cooling - the only cooling method that works within the lifetime of the muon - and emittance exchange, the desired emittances for a Higgs Factory or higher energy collider are attainable.  A cooling channel utilizing gas filled radio frequency cavities has been designed to deliver the requisite cool muon beam.  Technology development of these RF cavities has progressed from breakdown studies, through beam tests, to dielectric loaded and reentrant cavity designs.  The results of these experiments are summarized.}

\collaboration[c]{on behalf of the Muon Accelerator Program}

\begin{document}
\maketitle
\flushbottom

\section{Introduction}
\label{sec:Introduction}

A muon accelerator offers a diverse physics program, ranging from a high intensity Neutrino Factory to a high energy Muon Collider.  Such a machine has advantages over hadron and electron machines in terms of both physics reach and cost.  To realize a Neutrino Factory or to a larger degree a Muon Collider, a number of innovative techniques for accelerator subsystems must be proven to be feasible.  One such subsystem is the cooling channel, which is particularly important to achieving a luminosity on the order of 10$^{34}$ cm$^{-2}$s$^{-1}$ for modern colliders.  Approximately six orders of magnitude of cooling is needed because muon beams derived from colliding high power proton beams with a target are born with a large phase space.  Cooling channels for muon accelerators based on high pressure gas filled radio frequency cavities provide equilibrium emittances of 0.62 mm transversely and 0.89 mm longitudinally, close to the transverse design value, and better than the longitudinal design value for a Higgs Factory \cite{HCCEmittance}.  A research program whose purpose has been to validate the feasibility of high pressure gas filled radio frequency cavities has come to a conclusion under the auspices of the United States based Muon Accelerator Program.  The results of this program are outlined here.

\section{Ionization Cooling}
\label{sec:IonizationCooling}

The mechanism used to cool a beam of muons within their 2.2 $\mu$s lifetime is called ionization cooling, and works by passing the beam through an absorbing material and replacing the lost longitudinal energy \cite{IonizationCoolingBudker, CoolingSkrinskii, CoolingNeuffer}.  This is intrinsically a transverse effect, so to cool longitudinally, emittance exchange is introduced through dispersion in the beam.

The change in normalized transverse emittance with path length is given by:
\begin{equation}
\frac{\mathrm{d}\epsilon_{N}}{\mathrm{d}s} \approx -\frac{\epsilon_{N}}{\beta^{2}E} \frac{\mathrm{d}E}{\mathrm{d}s} +
\frac{\beta_{\perp}E_{s}^{2}}{2\beta^{3}m_{\mu}c^{2}L_{R}E}
\label{eq:Emittance}
\end{equation}
where $\beta = v/c$, $E$ is the beam energy, d$E$/d$s$ is the energy loss rate in the absorber, $\beta_{\perp}$ is the betatron function, $E_{s}$ is the characteristic scattering energy ($\approx$13.6 MeV), and $L_{R}$ is the radiation length of the absorber \cite{CoolingNeuffer1}.  The first term is the cooling term, and the second is heating due to multiple scattering in the absorber.

To maximize cooling, a large energy loss rate and radiation length, and small betatron function are required.  The energy loss rate and radiation length are determined by choice of the absorber material.  Hydrogen provides the most ideal combination, with $\mathrm{d}E/\mathrm{d}s\approx\,4.3\,\mathrm{MeV}\,\mathrm{cm}^{2}/\mathrm{g}$ for 200 MeV/c muons, and $L_{R}\,=\,63\,\mathrm{g}/\mathrm{cm}^{2}$.  To minimize the betatron function in the absorber, strong magnetic fields are required.

\section{Helical Cooling Channel}
\label{sec:HCC}

\begin{figure}[htbp]
\centering
\includegraphics[width=0.46\textwidth]{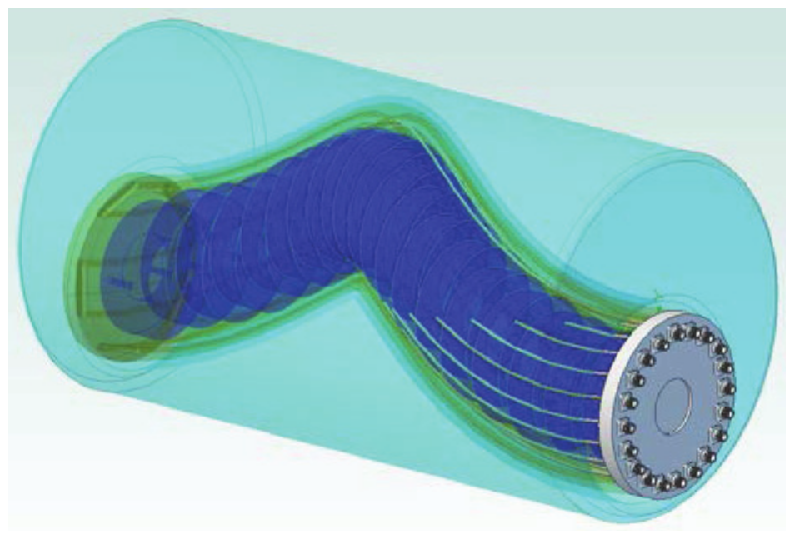}
\includegraphics[width=0.53\textwidth]{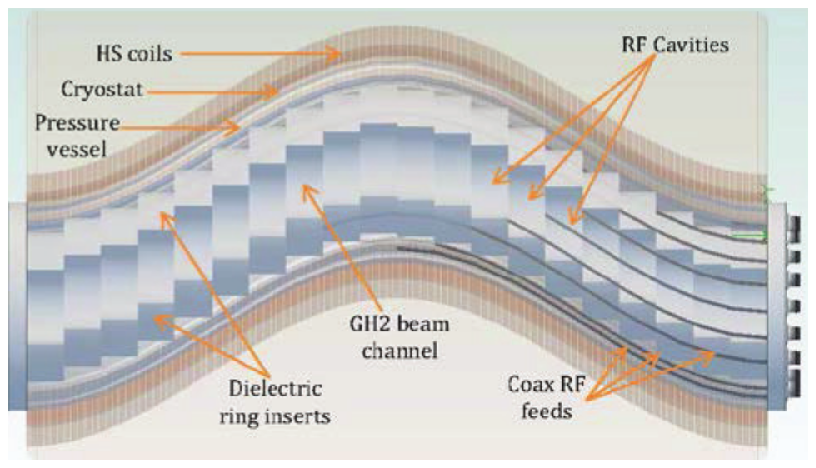}
\caption{Three dimensional view of one cooling cell of the Helical Cooling Channel (left) and cutaway view showing components (right) \cite{HCCJohnson}.}
\label{fig:HCC}
\end{figure}

One cooling channel scheme that has been studied extensively is called the Helical Cooling Channel (HCC) \cite{HCCDerbenev}.  The HCC provides continuous six dimensional cooling by placing RF cavities filled with high pressure hydrogen gas (the material needed for ionization cooling) along the helical trajectory of a beam traversing a magnetic field with solenoidal, helical dipole, and helical quadrupole components, thereby creating simultaneous energy loss and dispersion.  Figure \ref{fig:HCC} shows a three dimensional view of one cooling cell, as well as a cutaway view.

In this configuration, radio frequency (RF) cavities are subject to multi-tesla magnetic fields, dictating that the cavities be normal conducting.  Tests of normal conducting copper vacuum cavities showed significant degradation of the accelerating field when placed in an external solenoidal magnetic field \cite{BreakdownNorem, BreakdownMoretti}.  In addition to being the cooling medium, gas within a cavity provides the additional benefit of providing an energy loss mechanism for the dark current responsible for RF breakdown.  Provided free electrons do not gain enough energy to ionize a gas molecule, the gas should completely suppress breakdown.

\section{High Pressure Test Cell}
\label{sec:HPRFCavity}

\begin{figure}[htbp]
\centering
\includegraphics[width=0.42\textwidth]{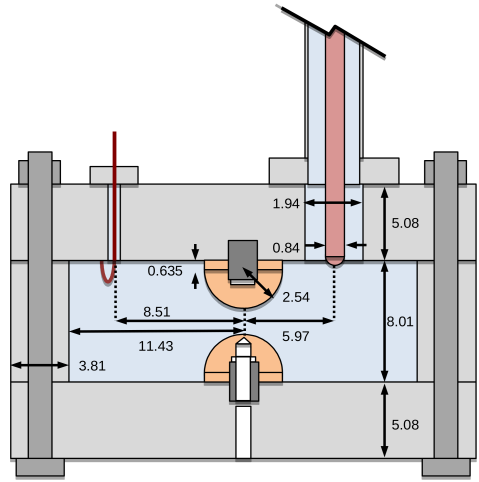}
\caption{Cutaway view of the HPRF cavity \cite{PRABFreemire}. The electrodes, RF coupler, and one RF pickup are shown, with the appropriate dimensions given.  Not shown are the gas feedthrough, one RF pickup, and six optical ports.}
\label{fig:HPRFCartoon}
\end{figure}

A copper coated stainless steel pillbox cavity was designed to test the breakdown limit for various gases and metals.  Removable hemispherical electrodes could be inserted at the center of each endplate to study the effect of metal on breakdown.  In the pillbox configuration with 1 atm air, the cavity was resonant at 1004.06 MHz, with an unloaded quality factor, Q$_{0}$, of 15,740. With the electrodes, the cavity was resonant at 800-810 MHz, with a Q$_{0}$ of 13,900-14,200, depending on gas species and pressure.  Figure \ref{fig:HPRFCartoon} shows a cutaway diagram of the high pressure test cell (herein referred to as the HPRF cavity).

The electrodes enhance the electric field on the cavity axis, ensuring breakdown occurs in a localized region.  One electrode and cavity endplate were also counterbored in order to minimize the material a proton beam would interact with before entering the cavity.

As a convention for reporting gas pressures was not adopted, it will be useful for the reader to recall that 1 atm = 14.7 psi = 760 torr when interpreting data in the following sections.

\section{Breakdown Tests}
\label{sec:Breakdown}

\begin{figure}[htbp]
\centering
\includegraphics[width=0.8\textwidth]{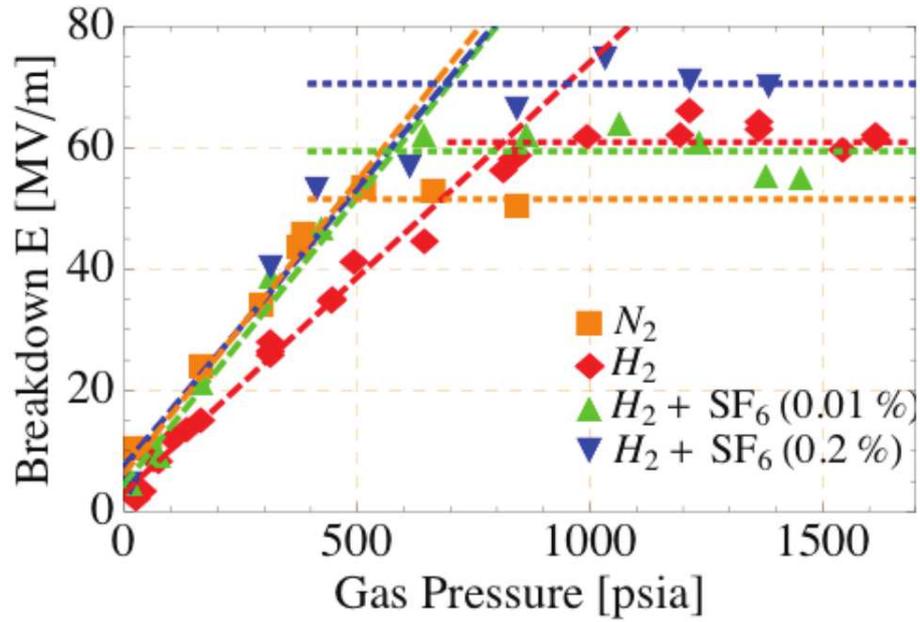}
\caption{Breakdown electric field vs gas pressure for pure nitrogen and hydrogen, and hydrogen doped with sulfur hexafluoride \cite{HPRFYonehara}.}
\label{fig:BreakdownGas}
\end{figure}

The RF breakdown gradient of hydrogen, nitrogen, and helium gas filled cavities, including cases with small concentrations of sulfur hexafluoride, were measured at the MuCool Test Area at Fermilab.  The results for nitrogen and hydrogen are shown in Figure \ref{fig:BreakdownGas}.  It can be seen that pure hydrogen and nitrogen follow the Paschen curve \cite{Paschen} until the point at which the gas can no longer suppress breakdown induced by the metal.  Gradients above 60 MV/m were attainable with pure hydrogen.

\begin{figure}[htbp]
\centering
\includegraphics[width=0.8\textwidth]{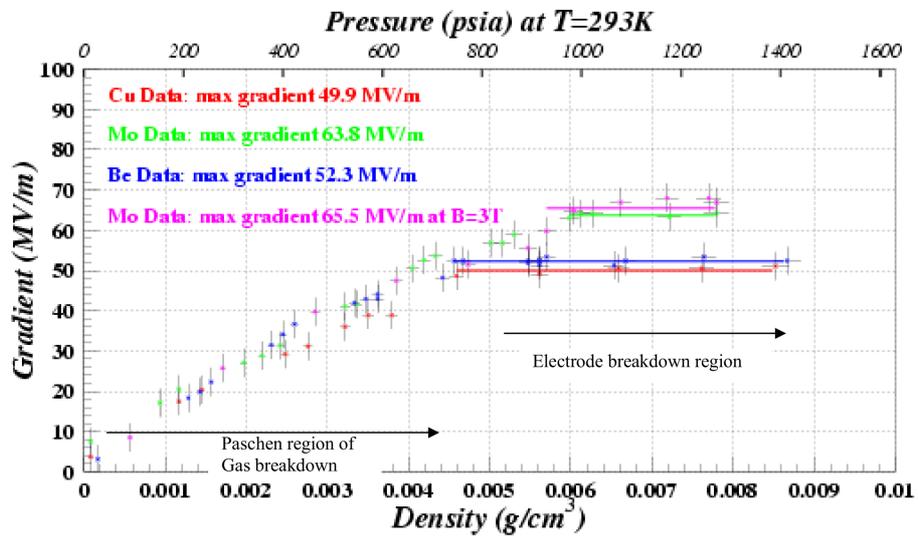}
\caption{Breakdown electric field vs hydrogen gas pressure for copper, molybdenum, and beryllium electrodes \cite{HPRFHanlet}.  Molybdenum data were collected with and without a 3 T external solenoid field.}
\label{fig:BreakdownMetal}
\end{figure}

The RF breakdown gradient of copper, molybdenum, and beryllium electrodes was also determined using hydrogen gas.  The results are shown in Figure \ref{fig:BreakdownMetal}.
The Paschen curve is observed for hydrogen, with the plateau in gradient dependent on electrode material.  The most important result is that the breakdown gradient is the same within the error bars for molybdenum data collected with and without an external 3 T solenoidal magnetic field.  This demonstrates that a high pressure gas can mitigate breakdown in RF cavities within a multi-tesla external magnetic field.

\section{Beam Tests}
\label{sec:BeamTest}

The HPRF cavity was subjected to a high intensity proton beam coming from the Fermilab linac in order to characterize the effect of beam on the cavity performance.  The maximum beam intensity was $2\times10^{8}$ protons per bunch for a 9.5$\,\mu$s long pulse bunched at 201 MHz.  The RF pulse length was 40$\,\mu$s and gradients up to 50 MV/m were run.  Data with and without a 3 T solenoidal magnetic field were recorded.

\begin{figure}[htbp]
\centering
\includegraphics[width=0.8\textwidth]{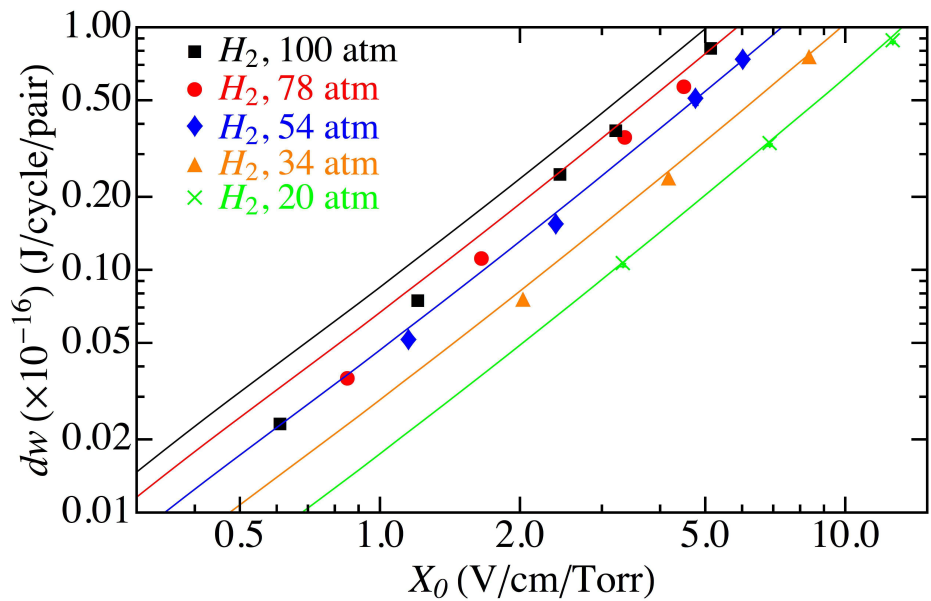}
\caption{Energy loss per electron-ion pair per RF cycle ($dw$) in the HPRF cavity vs the ratio of electric field to gas pressure ($X_{0}$) \cite{PRLChung}.  Various hydrogen gas pressures are plotted.  The curves are predictions for the energy loss based on electron drift velocity measurements in a DC electric field in low pressure hydrogen gas \cite{Lowke}.}
\label{fig:dw}
\end{figure}

Charged particles in the cavity gain energy from the electric field and transfer it to the gas, in a process called plasma loading.  The amount of plasma loading per electron-ion pair per RF cycle was measured for various values of the ratio of the electric field and gas pressure, and is shown in Figure \ref{fig:dw}.

\begin{figure}[htbp]
\centering
\includegraphics[width=0.8\textwidth]{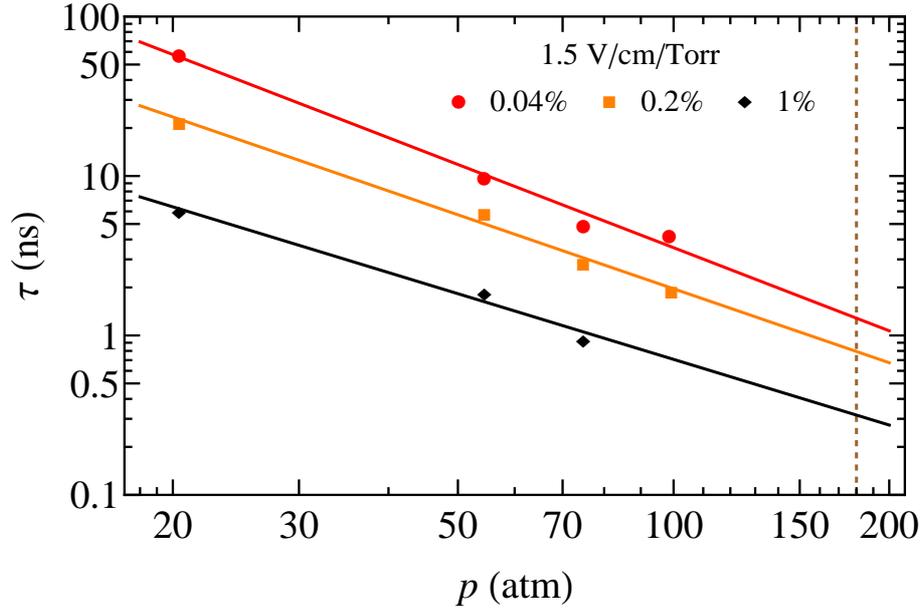}
\caption{Attachment time of electrons to oxygen molecules ($\tau$) vs hydrogen gas pressure for various concentrations of dry air at a fixed value of $X_{0}$ \cite{PRLChung}.  The solid lines are fits to the data, used to extrapolate the electron lifetime.  The vertical dashed line is 180 atm, a pressure under consideration for the Helical Cooling Channel.}
\label{fig:Tau}
\end{figure}

The rate at which electrons become attached to an electronegative dopant gas molecule was also measured, and is shown for the case of oxygen in Figure \ref{fig:Tau}.  Minimizing the electron lifetime minimizes plasma loading, which improves the performance of the cavity, and it can be seen that increasing the oxygen concentration decreases the attachment time.

\section{Simulation}
\label{sec:Simulation}

\begin{figure}[htbp]
\centering
\includegraphics[width=0.8\textwidth]{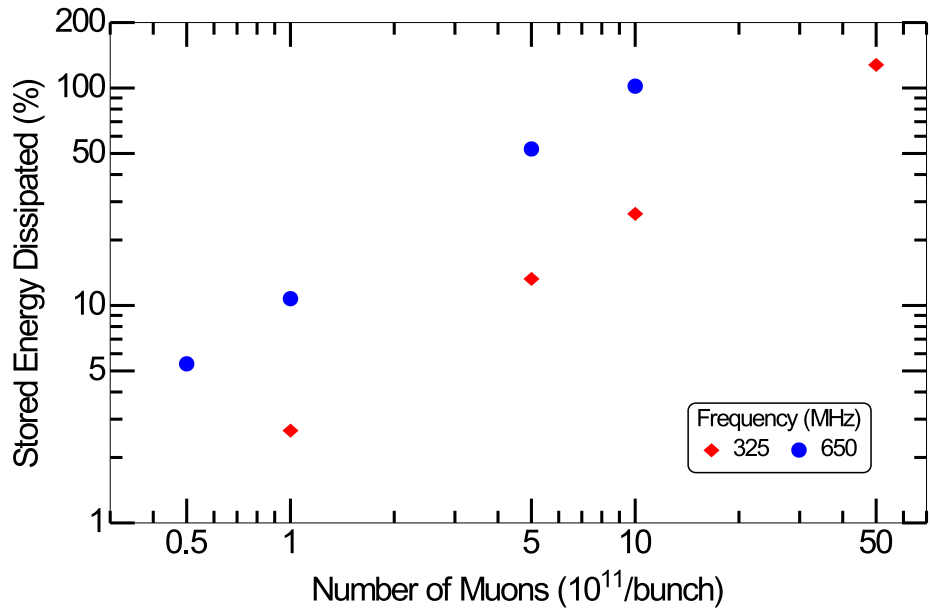}
\caption{Percent of the total stored energy for 325 and 650 MHz RF cavities considered for the HCC lost to plasma loading as seen by the final bunch in a 21 bunch train vs number of muons in each bunch \cite{PRABFreemire}.  There is no stored energy left in the cavity when 100\% dissipation is reached.}
\label{fig:PlasmaLoading}
\end{figure}

The results for the electron-ion energy loss and electron lifetime, as well as measurements of electron-ion and ion-ion recombination were used to predict the plasma loading in the Helical Cooling Channel \cite{PRABFreemire}.  Figure \ref{fig:PlasmaLoading} shows the percent of total stored energy in a 325 or 650 MHz HCC RF cavity lost to plasma loading for a bunch train of muons as a function of the number of muons in each bunch.  As the plasma created by each bunch dissipates a portion of the stored energy of the cavity, each subsequent bunch experiences a slightly smaller accelerating gradient.  This leads to a larger equilibrium emittance than would exist without plasma loading.  These predictions also give an indication of the maximum bunch population the HCC could accommodate.

\begin{figure}[htbp]
\centering
\includegraphics[width=0.8\textwidth]{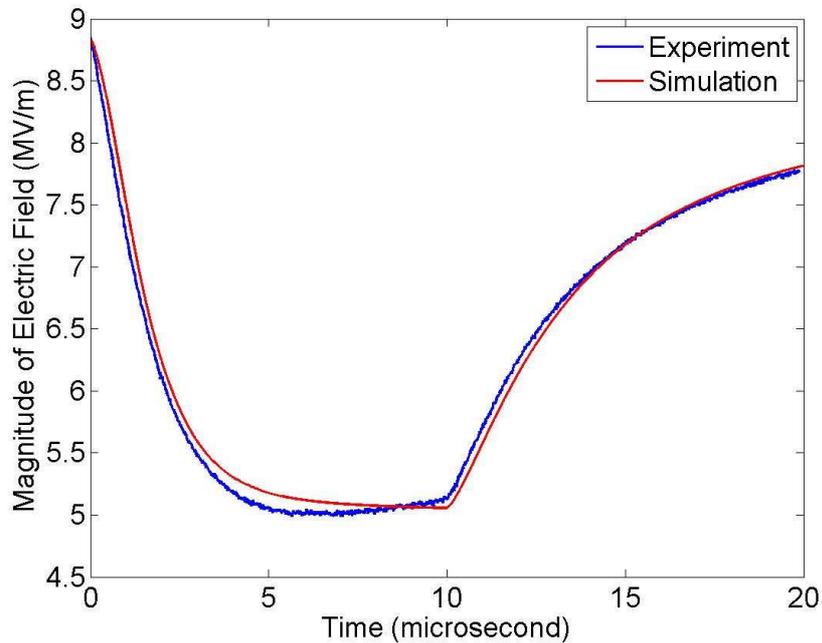}
\caption{Comparison between experimental data and simulation results using SPACE \cite{HPRFSimulation}.  The magnitude of the electric field vs time is plotted.  Time = 0 corresponds to the beam entering the cavity, with the beam pulse ending at 10 $\mu$s.  The electric field in the cavity recovers after the beam passes through.  (The turn on and turn off of the RF pulse have been omitted.)}
\label{fig:HPRFSimulation}
\end{figure}

A 3D electromagnetic particle-in-cell code with atomic physics processes, SPACE, has been developed to better predict plasma loading over a parameter range larger than was able to be recorded in the HPRF beam tests \cite{HPRFSimulation}.  SPACE contains many features relevant to HPRF cavities that aren't available in other codes, including: charge creation (ionization) and neutralization (recombination/attachment), varying time scales (from sub-nanosecond processes such as attachment, to microsecond RF and beam pulses), radio frequency electromagnetic fields, and an algorithm for moving between relativistic beam and laboratory frames containing individual particles.  The code has been benchmarked using results from the HPRF beam test for comparison.  Figure \ref{fig:HPRFSimulation} shows the agreement between the simulation code and experimental data for one parameter setting.  SPACE provides a predictive tool for investigating the total plasma loading using design parameters for a Helical Cooling Channel.

\section{Dielectric Loaded Tests}
\label{sec:Dielectric}

Engineering constraints on the bore size of the superconducting solenoid magnets necessary to provide the required magnetic field dictate transverse RF cavity dimensions smaller than conventional hydrogen filled pillboxes.  To shrink the diameter of the gas filled cavities, they can be loaded with a material with a larger dielectric constant, or made reentrant.  Alumina is an attractive dielectric material due to its small loss tangent and ease of manufacturing.  Low power tests of alumina and other materials obtained from various manufacturers were performed to select the best combination of dielectric constant and loss tangent \cite{DielectricConstant}.

\begin{figure}[htbp]
\centering
\includegraphics[width=0.98\textwidth]{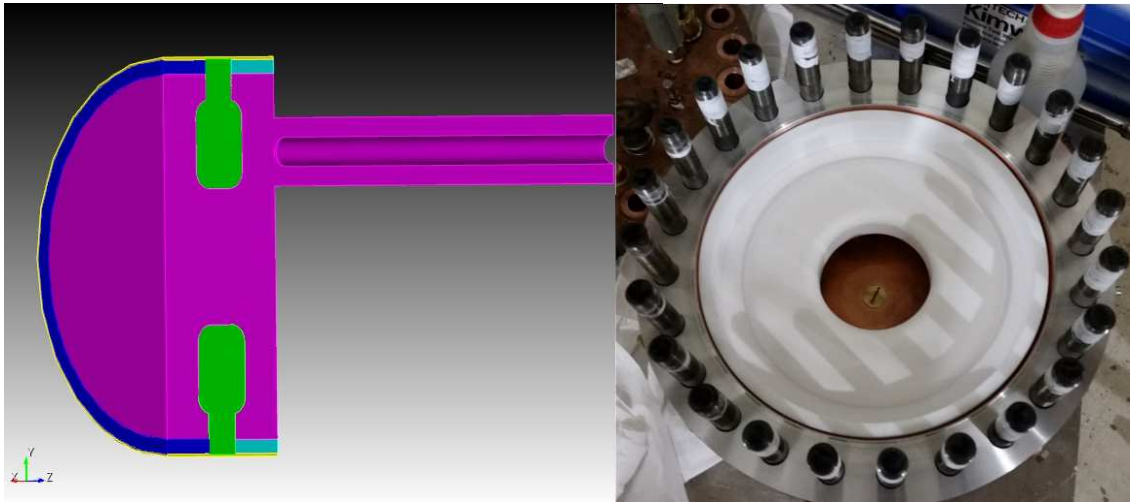}
\caption{Left - Cutaway view of the HPRF cavity (including coupler) with alumina insert (green) and teflon spacers (two shades of blue). The volumes shaded magenta and yellow were filled with nitrogen gas.  Right - Photo looking into the HPRF cavity with one endplate removed, showing the alumina insert and one spacer \cite{DielectricStrength}.}
\label{fig:DielectricInsert}
\end{figure}

Following the conclusion of the low power tests, four larger prototype alumina inserts of varying purity were fabricated for high power testing.  The HPRF cavity used for the breakdown and beam tests was modified by removing the electrodes to accommodate the frequency shift due to the dielectric.  Figure \ref{fig:DielectricInsert} shows a cutaway of the HPRF cavity with the alumina insert and teflon spacers and a photo of the cavity with one endplate removed, exposing the alumina and one spacer.

\begin{figure}[htbp]
\centering
\includegraphics[width=0.48\textwidth]{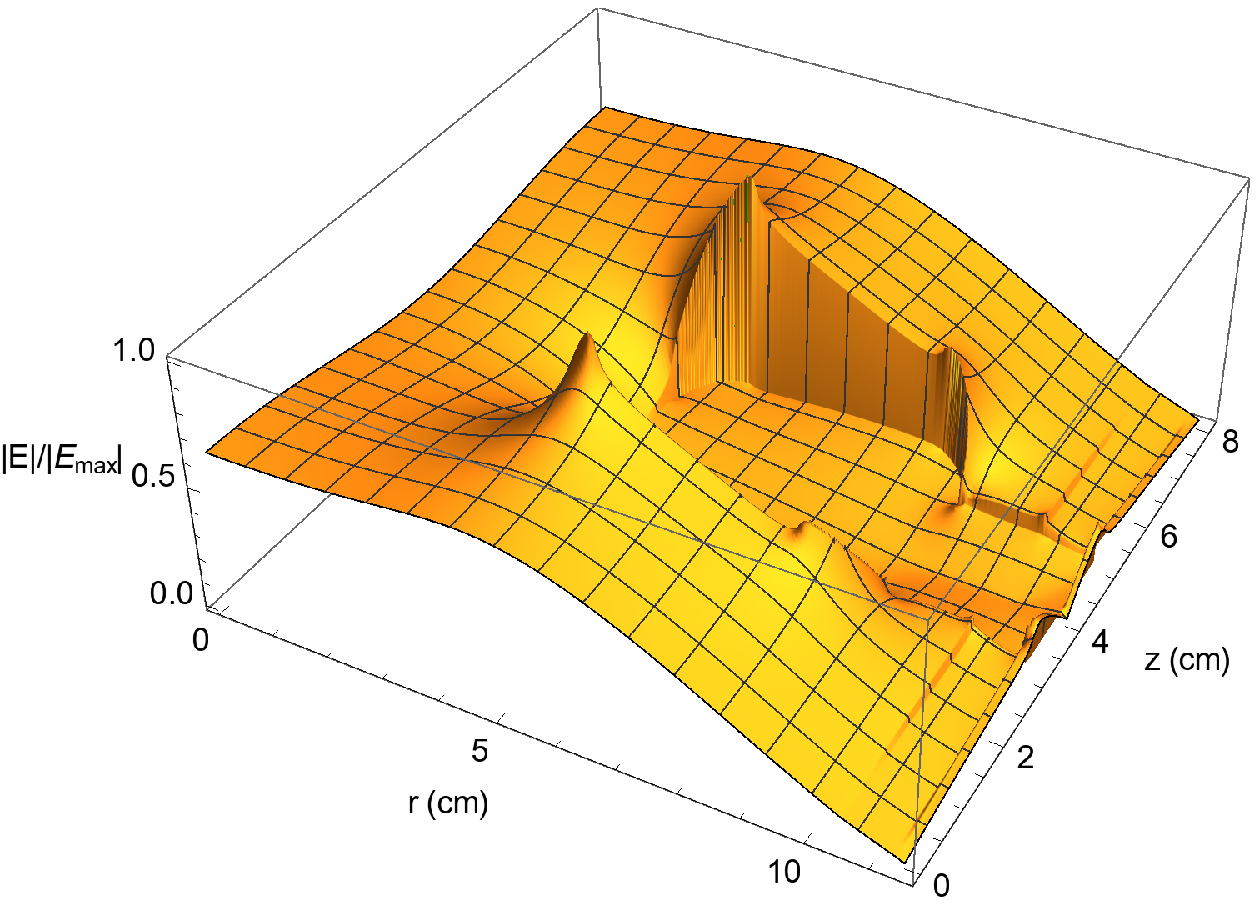}
\includegraphics[width=0.48\textwidth]{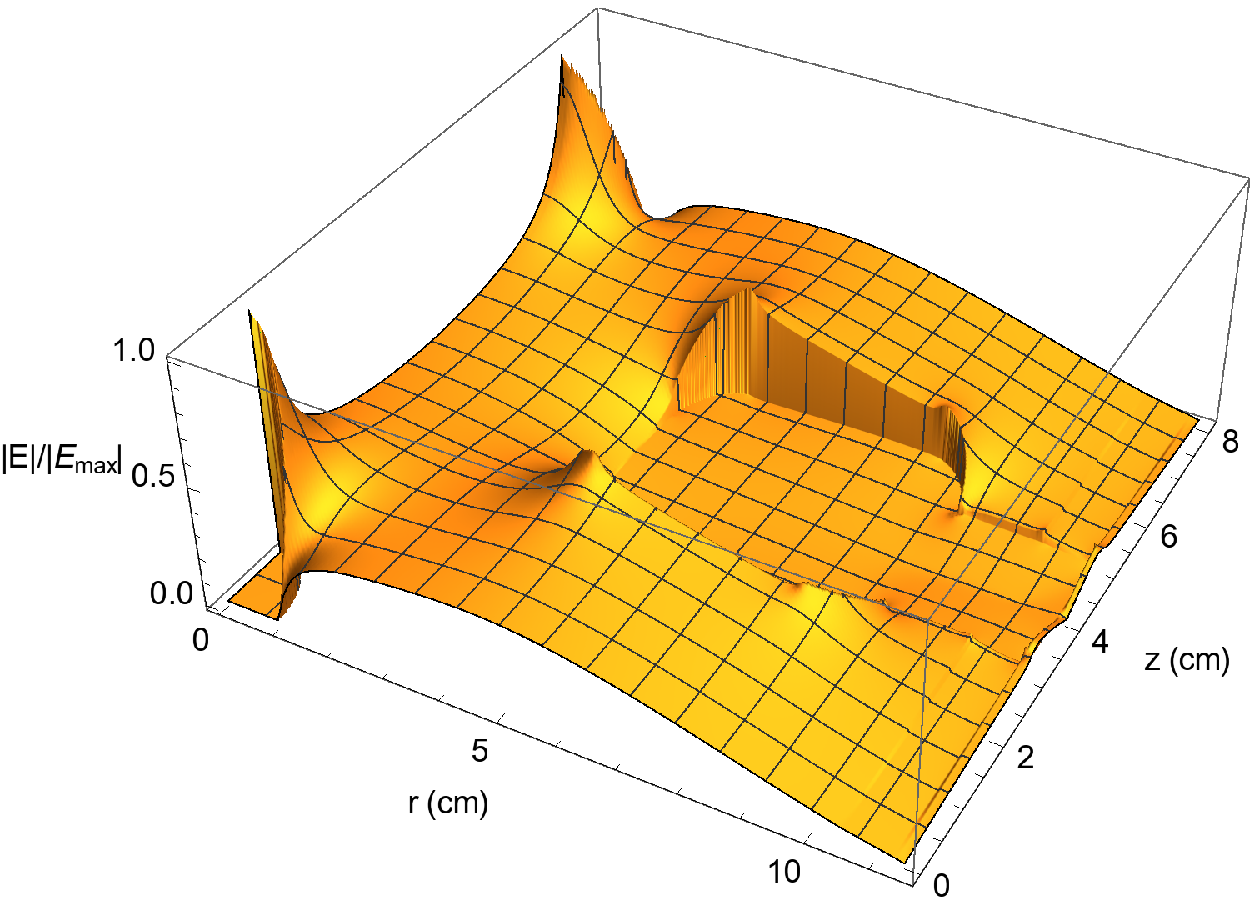}
\caption{Normalized electric fieldmaps of the HPRF cavity with alumina insert and spacers in two configurations:  no electrodes (left), and with two 0.95 cm radius electrodes on axis (right) \cite{DielectricStrength}.}
\label{fig:DielectricFieldmaps}
\end{figure}

The geometry of this design ensured the triple point junction (the place at which metal, ceramic, and gas meet) was in a region of low electric field, and minimized the volume of dielectric needed by placing the alumina in a region of large electric field.  The electric fieldmap of the cavity in this configuration is shown in Figure \ref{fig:DielectricFieldmaps} (left).

\begin{figure}[htbp]
\centering
\includegraphics[width=0.8\textwidth]{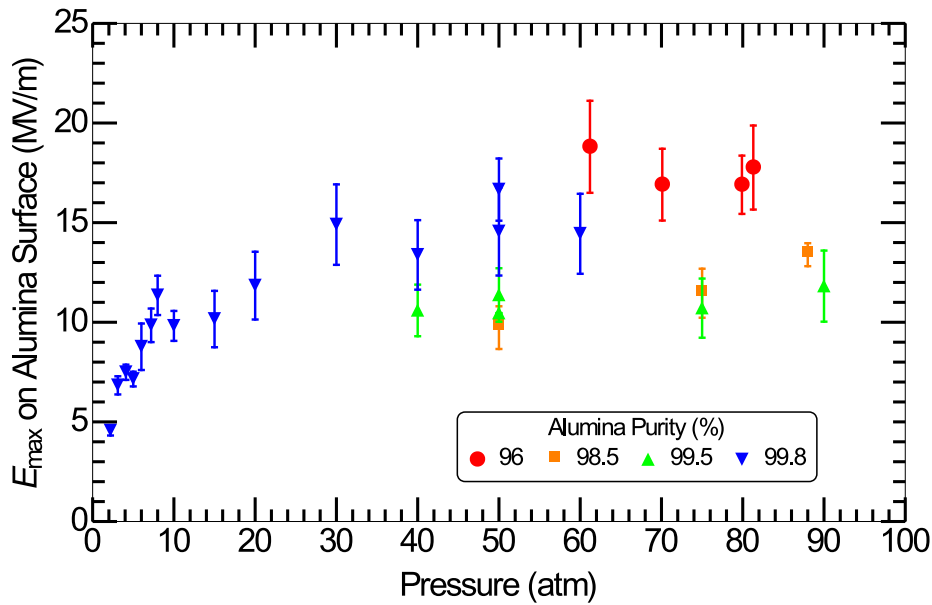}
\caption{Peak electric field on the surface of the alumina insert at which sparking was observed as a function of nitrogen gas pressure for alumina purities of 96, 98.5, 99.5, and 99.8\% \cite{DielectricStrength}.}
\label{fig:DielectricBreakdown}
\end{figure}

It can be seen from Figure \ref{fig:DielectricFieldmaps} (left) that the peak electric field is on the interior rounded part of the alumina insert.  The peak surface field on the alumina at which the cavity sparked was measured for alumina purities ranging from 96 to 99.8\% as a function of nitrogen gas pressure \cite{DielectricStrength}.  The results are shown in Figure \ref{fig:DielectricBreakdown}.
The Paschen curve for nitrogen was observed at pressures up to 10 atm, at which point the alumina limited the electric field within the cavity.  The highest purity alumina did not exhibit a significantly smaller dielectric strength, and is advantageous because of its smaller loss tangent.

\begin{figure}[htbp]
\centering
\includegraphics[width=0.8\textwidth]{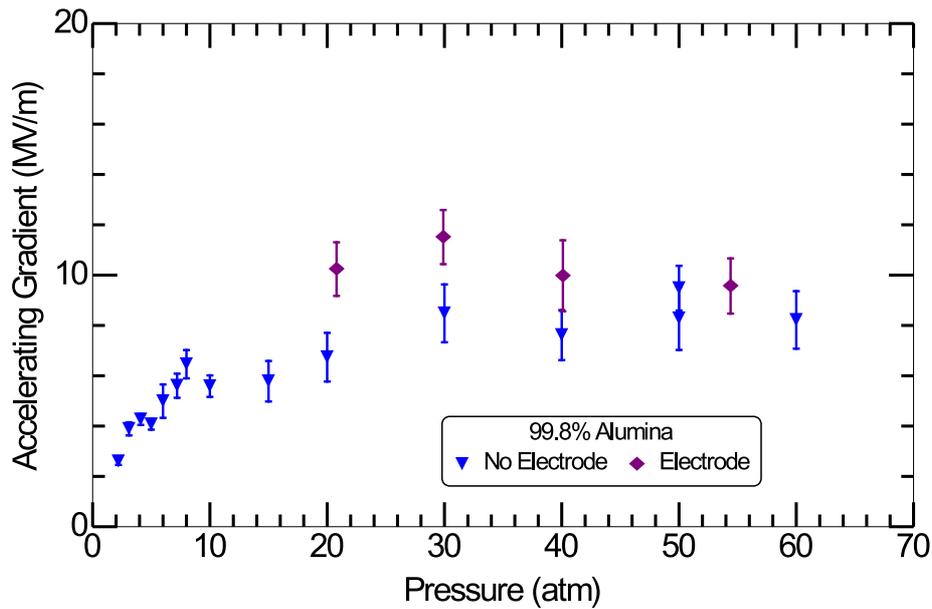}
\caption{Accelerating gradient at which sparking was observed for the 99.8\% alumina insert with and without copper electrodes on axis \cite{DielectricStrength}.}
\label{fig:DielectricBreakdownElectrode}
\end{figure}

In an attempt to increase the accelerating gradient achievable within the cavity with the 99.8\% alumina insert, small electrodes were fabricated and inserted on the cavity axis.  This small perturbation in the electric field shifted the location of the peak electric field to the tip of the electrodes, as shown in Figure \ref{fig:DielectricFieldmaps} (right).  Measurements of the gradient at which the cavity sparked in this configuration showed the peak field on the alumina surface remained constant, while the accelerating gradient of the cavity improved by 23\% on average, as shown in Figure \ref{fig:DielectricBreakdownElectrode}.

Further optimization of cavity and insert geometry is necessary in order to achieve a larger accelerating gradient while minimizing power losses.

\section{Conclusion}
\label{sec:Conclusion}

A research and development program spanning more than a decade aimed at demonstrating the feasibility of high pressure gas filled radio frequency cavities for use in muon cooling channels has come to a conclusion.  The technology validation of HPRF cavities has progressed from RF breakdown measurements, through beam-induced plasma loading tests, to engineering studies based on dielectric loading.  The results indicate that gas filled RF cavities could be designed to meet the performance requirements for a muon cooling channel.

\acknowledgments

This work has been made possible through support from Fermilab, operated by Fermi Research Alliance, LLC under Contract No. DE-AC02-07CH11359 with the U.S. Department of Energy, and funding from the United States Department of Energy Small Business Innovation Research (SBIR) and Small Business Technology Transfer Research (STTR) grant programs.


\end{document}